\documentclass[aps,pre,twocolumn,groupedaddress,showpacs]{revtex4}
\usepackage{latexsym}
\usepackage{times}
\usepackage{graphicx}
\usepackage[tight,FIGTOPCAP]{subfigure}
\newcommand{\be}{\begin{equation}}
\newcommand{\ee}{\end{equation}} 

\begin{document}
\title{Sedimentation of pairs of hydrodynamically interacting
  semiflexible filaments } 
\author{I. Llopis}

\author{ I.  Pagonabarraga} 
\affiliation{Departament de F\'{\i}sica
  Fonamental, Universitat de Barcelona, C. Mart\'{\i} i Franqu\'es 1, 08028
  Barcelona, Spain} 

\author{M. Cosentino Lagomarsino} 
\affiliation{UMR 168 / Institut Curie, 26 rue d'Ulm 75005 Paris, France}
\affiliation{Universit\`a degli Studi di Milano, Dip.
    Fisica, and I.N.F.N., Via Celoria 16, 20133 Milano, Italy,  } 
\author{C.P. Lowe}
\affiliation{Van't Hoff Institute for Molecular Science, University of
  Amsterdam, Nieuwe Achtergracht 188, 1018 WV Amsterdam, The Netherlands} 

\date{\today}

\begin{abstract}
  We describe the effect of hydrodynamic interactions   in the
  sedimentation of a pair of inextensible semiflexible filaments  under a
  uniform constant force at low Reynolds numbers. We have
  analyzed the different regimes and the morphology of such polymers in
  simple geometries, which allow us to highlight the peculiarities of the interplay between elastic and hydrodynamic stresses.  Cooperative and symmetry breaking effects associated to the geometry of the fibers gives rise to characteristic motion which give them distinct properties from rigid and elastic filaments.
\end{abstract}
%\PACS{05.10.-a,05.65.+b}
\maketitle
%%%%%%%%%%%%%%%%%%%%%%%%%%%%%%%%
\section{Introduction}
%%%%%%%%%%%%%%%%%%%%%%%%%%%%%%%%

The understanding of the hydrodynamics of semiflexible mesoscopic
filaments has gained interest due to the relevance of these fibers in
different contexts. Many biopolymers are virtually inextensible
semiflexible and their dynamics in a fluid plays a central role in the
motion of cilia, and eukaryotic and prokaryotic
flagella~\cite{alberts_cell}. Although cell motility has been
investigated for decades~\cite{bray}, recent advances in
microfabrication and micromanipulation enable us to interact directly
with them in simplified \emph{in vitro} environments, where
most of the parameters are under control. This allows direct and well defined 
measurements.
For example, Riveline and coworkers~\cite{riveline} have employed optical
tweezers to periodically oscillate actin filaments connected to
micron-sized beads, in order to devise an  artificial ``one-armed
swimmer''.  More recently, Dreyfus {\sl et al}~\cite{Dreyfus} have  been able to produce artificial swimmers out of polymer-linked magnetic beads. This approach  enables an easier control of the filaments through magnetic fields, and has allowed to perform quantitative measurements of the physical properties of the chains, such as their bending stiffness, opening a new method to induce properties of the linker molecules~\cite{gobault} or the affinity
of the chemical contacts between polymer and particle coating from
simple mesoscopic measurements, such as image analysis from
video-microscopy~\cite{koenig}.
Many other possibilities remain  to be explored, ranging from the use of
semiflexible filaments in microfluidic devices to the fabrication of synthetic
ciliary arrays, or to technological applications of artificial swimmers.

%% theoretical motivation
These advances give a renewed stimulus to a quantitative and careful
analysis of the hydrodynamics of semiflexible
filaments~\cite{wiggins_bio,marco2,roper,cebers}.  They differ from
flexible polymers in the ways in which elastic and hydrodynamic
stresses compete, and it is necessary to treat both on the same
footing, a theoretical challenge.
To gain understanding in this interplay and its dynamic implications it
is useful to consider the motion of semiflexible chains subject to a
uniform external driving. A single sedimenting filament has been
considered recently, and it has been shown
theoretically~\cite{marco3,xunadim,Netz_Sed} that the chain response
differs qualitatively from that of a rigid rod, in accordance with
predictions on collective properties of fiber
suspensions~\cite{butler06}. Specifically, the inhomogeneous hydrodynamic
stress along the fiber induced by hydrodynamic interactions (HI) leads to
filament bending and orientation transverse to the applied
field~\cite{marco3}. Upon increasing the driving, the shape of the
filament changes and becomes eventually unstable; the filament then
sediments without reaching a steady state.

We will analyze the interactions between a pair of sedimenting
filaments, and will study how the combined effect of HI and elasticity
induce cooperativity in their motion. The response of the fibers
depends on the specific geometry; in particular the
translation-rotation coupling is sensitive to the symmetry of the
relative positioning of the fibers. Although it is known that
translation-rotation couplings lead to an intricate behavior in the
sedimentation of rigid rod suspensions~\cite{butler02}, flexibility
leads to new scenarios.

Hence, we will consider three simplified situations. In order to address
the role of hydrodynamic cooperativity in the absence of symmetry breaking
and induced rotation, we will analyze first two cases where such a coupling
is prevented.  We are then able to find the relevant scaling regimes for
the velocity and the short- and long-time deformation amplitudes, as a
function of the inter-filament distance.
Subsequently, we will focus in two collinear filaments, the simplest geometry
where rotation is induced.  The proposed situations can be realized
experimentally in a straightforward way, to test our predictions.

The remainder of the paper is organized as follows.  In section~\ref{s2}
we present our computational model, defining the relevant parameters of
the system. In section~\ref{s3} we summarize the relevant features of
single filament sedimentation, which will be useful in subsequent
sections.  In section~\ref{s4}, we analyze the sedimentation of a pair
of semiflexible filaments.  We conclude in section~\ref{s8} with a
discussion of the main results and their implications.

%%%%%%%%%%%%%%%%%%%%%%%%%%%%%%%%
\section{Model}
\label{s2}
%%%%%%%%%%%%%%%%%%%%%%%%%%%%%%%%

We study numerically the dynamics of inextensible semiflexible filaments
of length $L$, which are characterized by their bending energy.  Such an approach is relevant for a large class of biological and non-biological
polymers including DNA, cytoskeletal filaments and carbon
nanotubes~\cite{Liverpool}, as well as for filaments where the degree of
extensibility is negligibly small.
 
A semiflexible filament can be described by the arclength distance along
the filament at a given time $t$, ${\bf r}(s,t)$, where $s \in
[0,1]$. Accordingly, its local curvature is $C[{\bf r}(s,t)] = |
\partial^2 {\bf r}(s,t) \ / \partial s^2|$, and once the inextensibility
constraint is enforced, the elastic energy is given by the Hamiltonian
\be {\mathcal H} = \frac{1}{2} \kappa \int_0^L
C[{\bf r}(s,t)]^{2} ds \ \ ,
\label{energy}
\ee
where $\kappa$  stands for the filament's stiffness. 

We model such a filament as a chain of $N$ spheres (``beads'') of radius
$a$, connected by bonds of fixed distance $b$. Correspondingly, the
bending energy is expressed as the discretization of the Hamiltonian,
\be
{\mathcal H}_b = \frac{\kappa}{b}
\sum_{i=2}^{N-1} \left(1 - \cos \theta_i \right)  \ \ ,
\label{eq:energy}
\ee
where $\theta_i$ is the angle between the bond that connects bead $i-1$
to bead $i$ with the one that connects bead $i$ with bead $i+1$. The
need to resolve the conformational change of the filament makes the
simulations computationally much more intensive than when considering
rigid rods~\cite{butler02}.
The change in bending energy ${\mathcal H}_b$ due to the change in
position of bead $i$ determines the bending restoring force acting on
it, ${\bf F}_{iB}$. Inextensibility implies that the total polymer
length is fixed, $L=b (N-1)$; this quantity is kept constant by
constraint forces, ${\bf F}_{iC}$, applied at every time step on each
bead $i$~\cite{marco2,Lowe}.

It is usual to find biopolymers in suspension.  Accordingly, we need to
account for the interactions with the embeding solvent. Since the
Reynolds numbers are small (in water suspensions, $Re\sim 10^{-6}$ for
micron size filaments moving at micrometer per second, and $Re\sim 1$
only for millimeter size filaments displacing at millimeter per second)
we need to account for the coupling between elastic and viscous stresses
acting on the chains. To this end, we consider that each bead $i$ is
subject to a local friction force
\be
{\bf F}_{iF} = -\gamma_0 ({\bf v}_i - {\bf v}_i^H),
\label{eq:friction}
\ee
where $\gamma_0=6\pi\eta a$ is a friction coefficient related to the bead size and the solvent viscosity, $\eta$,
while $ {\bf v}_i^H$ stands for the velocity of the solvent generated by
the forces that the rest of the beads exert on the fluid at the position
of bead $i$. These dissipative forces couple hydrodynamically all the
chain beads through the solvent, giving rise to the hydrodynamic
interactions (HI). According to the standard procedure in polymer
physics, we describe the flow generated by the filament at the level of
the Oseen approximation. Hence, the induced velocity at the position of
bead $i$ can be expressed as~\cite{doi, marco3}
\be
{\bf v}_i^H (t) = \frac{3}{4 \gamma_0} \frac{a}{b}  \sum_{j \neq  i} \frac{1 + {\bf e}_{ij} (t) {\bf e}_{ij} (t)}{r_{ij} (t) / b}
\cdot {\bf F}_{j} (t),
\label{Oseen}
\ee
where $a$ determines the hydrodynamic coupling relevance while ${\bf
e}_{ij} = ({\bf r}_i - {\bf r}_j)/ r_{ij}$ is the unit vector joining
beads $i$ and $j$,with $r_{ij} = |{\bf r}_i - {\bf r}_j|$ being the
distance between them; $ {\bf F}_{j}$ refers to the total force acting
on bead $j$. Although alternative and more accurate approximations to
the induced velocities can be implemented~\cite{brenner}, this simple
coupling is enough to capture the essentials of HI on elastic
filaments, although for small filament separations our prediction will not be in general quantitatively accurate. Following the usual approach in polymer physics, we take
$b=2a$, consistent with the Shish-kebab model. The friction coefficient $\gamma_0
b/a$   provides a means to relate
the model parameters to physical units.  The local friction force gives
rise to an effective filament friction coefficient which depends on
filament configuration; therefore, this approach goes beyond resistive
force theory~\cite{brokaw,wiggins_bio}, which regards the solvent as a
passive medium that exerts a constant friction coefficient and which
does not account for the change in friction with filament configuration.

The description in terms of a local friction force allows to describe
the filament's dynamics using a molecular dynamics approach based on the
total force acting on each bead, ${\bf F}_i={\bf F}_{iB}+{\bf
F}_{iC}+{\bf F}_{iF}+{\bf F}_{ie}+{\bf F}_{ith}$, with ${\bf F}_{ie}$
and ${\bf F}_{ith}$ the external force and the random force which
accounts for thermal fluctuations respectively. Multiple filaments can
be also analyzed without any further algorithmic complexity. In this
paper we will concentrate on filament sedimentation, where ${\bf F}_{ie}
\equiv {\bf F}_e$ is a constant external force~\cite{Llopis_jnnf}, in
situations where the energetic contributions due to the elastic energy
dominate and will therefore neglect thermal forces (${\bf F}_{ith} =
{\bf 0}$).  Non-uniform oscilatory external drivings have also been
considered within the same approach in the context of filament
swimming~\cite{Lowe,marco2,Llopis_fla,Llopis_jnnf}.

Note that the choice $b=2a$ implies that, as the number of beads
increases, the hydrodynamic aspect ratio of the filament, $a/L$,
decreases as $1/N$. The basic hydrodynamic coupling we analyze depends
only marginally on this choice. A quantitatively more accurate
description of the filament finite thickness would require a
computationally costly description of the filament morphology,
and would not affect qualitatively the results. We have checked that if
we change the number of beads , $N$, but keep the mass and length of the
filament constant, the results obtained change less than a few percent
in the worst case for values ranging between $N=30$ and
$N=100$. Moreover, we observe a clear tendency to convergence on
increasing $N$ which implies that the trends described subsequently are
robust. The results we will describe have been obtained for filaments
with $N=30$. We have taken $L=1$ and the filament density $\rho=1$ while
the friction coefficient $\gamma_0=5$ sets the time scale. The bead mass
is then $m_b=\pi/2N^3$. Numerically, one has to resolve the non-physical
inertial time scale in which the bead's acceleration decays; hence we
take the time step $10^{-6}$. The bending rigidity was varied to control
the filament flexibility, but the characteristic associated time scale  is always larger than the
inertial time scale, ensuring that inertia becomes irrelevant at the
scales in which the filament configuration evolves.

%%%%%%%%%%%%%%%%%%%%%%%%%%%%%%%%
\section{Sedimentation of a single semi-flexible filament}
\label{s3}
%%%%%%%%%%%%%%%%%%%%%%%%%%%%%%%%

We briefly describe the main features of single filament sedimentation,
which has been explored both analytically and
computationally~\cite{marco3,Netz_Sed,xunadim}, in order to help with
the analysis in the coming sections. Filament sedimentation can be
described in terms of the dimensionless parameter $B = L^3 F_e /
\kappa$, the ratio of the energy imparted by  the external force and bending
energy~\cite{marco3}. We disregard thermal effects, which we consider
subdominant.

\begin{figure}
  \includegraphics[width=9cm,angle=0]{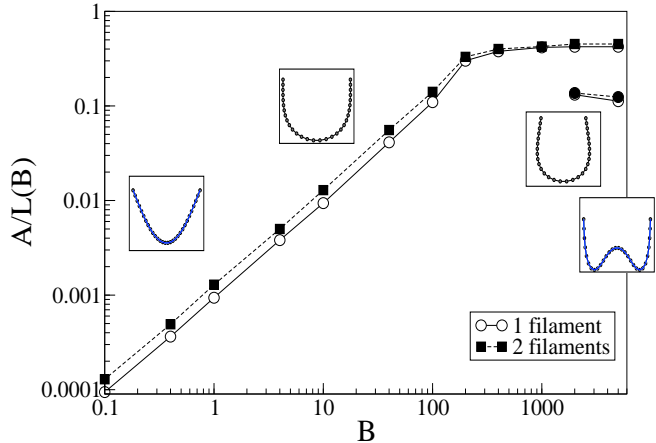} 
\caption{Bending amplitude of a
single and a pair of sedimenting filaments, separated a distance
$d/L=0.1$, as a function of the dimensionless parameter $B$.}
\label{fig:onefilament}
\end{figure}

If an external homogeneous force (${\bf F_e}$) is applied transversally
to the filament axis, its shape reaches a steady state as a result of
the competition between elastic, constraint, external and friction
forces. Since the friction force acting on beads near the chain's ends
is smaller than the local friction in their center, filaments bend and
align perpendicular to the externally applied field. For low to
intermediate values of B, chain sedimentation can be characterized by
the bending amplitude, $A$, defined as the distance between the highest
and lowest beads along the direction of the applied field.  For $B<50$,
the filament's amplitude increases linearly; at $B\sim 200 $ a plateau
is reached, signaling a saturation of the filament deformation in
response to the applied field, as depicted in
Fig.~\ref{fig:onefilament}. At even larger values of $B$, metastable
shapes with two minima are observed~\cite{marco3}.

If the chain is not aligned perpendicular to the applied field, the
friction force is not balanced and it generates a net torque that will
align the filament perpendicular to the external
force~\cite{marco3,xunadim}. For weak forcings, where the degree of
bending is proportional to $B$, the torque generated is also
proportional to $B$.
However, the time it takes the polymer to rotate increases as $1/B$ for
weak forcings, leading to a singular behavior for a rigid rod ($B=0$),
in which the filament keeps its initial orientation because it takes an
infinite time to rotate the filament to its perpendicular orientation, a
feature that is not captured by resistive force theory.
Hence, a single elastic polymer reacts to an applied field in a
qualitatively different way than a rigid one.
%%%%%%%%%%%%%%%%%%%%%%%%%%%%%%%%
\section{Sedimentation of a pair of flaments}\label{s4}
\begin{figure*}
  \includegraphics[width=18cm]{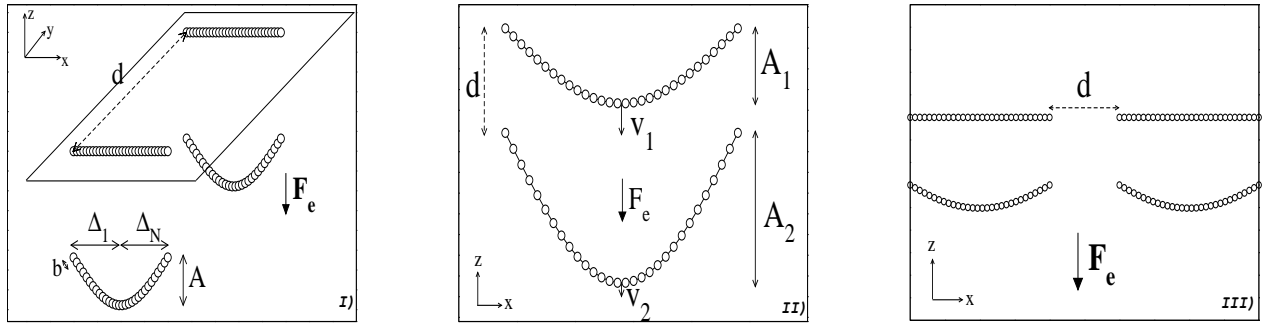}
  \caption{Pairs of semiflexible filaments characterized by their
    stiffness $\kappa$ and length $L$ that are separated a distance
    $d$, when they sediment under the action of a uniform external
    force field, $F_e$, for three different geometries; in the
    three cases we show the relevant parameters. I) Geometry I: Parallel filament sedimenting due to a uniform force  transverse to the plane defined by them. II) Geometry
    II: Sedimenting coplanar filaments. III) Geometry III: Sedimenting collinear
    filaments }\label{confi_ini_gen}
\end{figure*}

%%%%%%%%%%%%%%%%%%%%%%%%%%%%%%%%
%
We will now consider pairs of symmetric filaments of length $L$ and
rigidity $\kappa$, at a distance $d$ and subject to an external uniform
force field ${\bf F}_e$. We assume that ${\bf F}_e$ is parallel to ${\bf
e}_z$ and that the polymers lie initially perpendicular to the applied
field, along the ${\bf e}_x$ direction.
The details of the cooperativity induced by the hydrodynamic coupling
are sensitive to the initial configuration. To distinguish between
different effects induced by hydrodynamics, we will consider three
different geometries, as depicted in Fig.~\ref{confi_ini_gen}, which
correspond to parallel (geometries I and II), and collinear (geometry
III) filaments.
Geometries I (Fig.~\ref{confi_ini_gen}.a) and II
(Fig.~\ref{confi_ini_gen}.b) preserve the mirror symmetry with respect
to the filament's center while geometry III (Fig.~\ref{confi_ini_gen}.c)
will allow us to explore the effects of translation-rotation coupling. We will see
that geometry I conserves the symmetries of the one-filament case, and
geometry II breaks the up- down symmetry.

The presence of a second chain modifies the friction exerted on the
first filament. As a result, the filament shape and velocity will change
as a function of the distance between chains. Depending on their initial
conditions, the presence of a second thread can induce rotation of the
sedimenting polymer, breaking the mirror symmetry. We will characterize
this translation-rotation coupling through the deformation asymetry
parameter, \be D =\frac{\Delta_1 - \Delta_N}{\Delta_1 + \Delta_N}
\label{def:D} \ee
where $\Delta_k=\left\vert x_k - x_{min}\right\vert$ is the distance
along the $x$-direction between the $k$-th bead and the lowest bead, as
shown in Fig.~\ref{confi_ini_gen}.a. The parameter $D$ ranges between
$[-1,1]$, and reflects the transverse asymmetry of the filament
ends. For a single filament, there is no shape asymmetry and $D=0$.

%%%%%%%%%%%%%%%%%%%%%%%%%%%%%%%%
\subsection{Geometry I. Parallel Filaments Under a Force Transverse to
  the  Plane They Define. }
\label{s5}
%%%%%%%%%%%%%%%%%%%%%%%%%%%%%%%%

The first geometry under consideration involves two filaments that are
parallel and transverse to the external force
(Fig.~\ref{confi_ini_gen}.a).  Due to the initial configuration, they
will sediment at the same speed with $D=0$ and keeping their initial
separation $d$.
After a short time interval, in which HI propagate and the filaments'
inertia decays, they reach their steady state sedimentation velocity and
deform into shapes analogous to the ones described for single polymer
sedimentation.

\begin{figure*}[htp!]
\begin{center}
\includegraphics[width=16cm,angle=0]{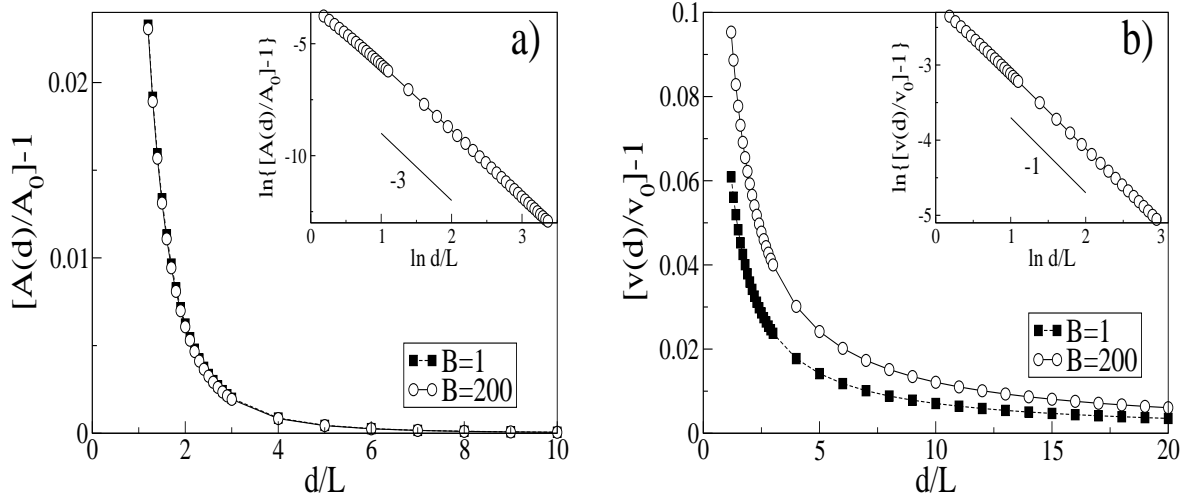}
\end{center}
\caption{a) Bending amplitude, $A$, and b) velocity in the direction of
  the force field as a function of filaments' separation for
  Geometry I, at $B=1$ and $B=200$. The asymptotic values of $A$ are
  defined as $A_0 \equiv A(d \to \infty)$, are $A_0(B=1)/L=9.43 \times
  10^{-4}$ and $A_0(B=200)/L=3.40 \times 10^{-1}$. Velocities are
  expressed in units of the sedimentation velocity of an isolated
  filament, $v_0(B)\equiv v(d \to \infty,B)$.  } \label{g1}
\end{figure*}

Since in our model HI propagate instantaneously, the sedimentation
velocity of the initially straight filaments after one time step will
deviate from its free draining value $v_0 \equiv
F_e/N\gamma_0$; in appendix~\ref{oseen_prediction} we compute this initial velocity. Subsequently, the filaments will deform until they reach
a new steady state in which they fall down at a different speed. Hence,
hydrodynamic cooperativity shows up in the degree of filament bending
and its dependence on chains separation.
\paragraph*{Bending amplitude.}
The filament deformation can be characterized through the same bending
amplitude, $A$, defined for single filament sedimentation.  For a given
value of $B$, $A$ will now depend on filaments' separation, $d$.  We
have found that $A(d)$ is consistent with an algebraic decay,
\be
A(d) \sim d^{-3}.
\ee
as displayed in Fig.~\ref{g1}.a. The dependence of $A$ on $B$ for a pair
of filaments separated a distance $d$ does not differ quantitatively
from the one observed for a single filament if we compare the filament's
shape with equal values of the parameter $A$. The effect of the second
filament can hence be understood in terms of an effective bending
energy. Making use of Fig.~\ref{fig:onefilament}, it is possible to
reproduce the filament's shape by identifying $B_{eff}(d)$ once $A$ has
been measured.
\paragraph*{Sedimentation velocity.}
The presence of a second filament leads to an increase of the
sedimentation velocity, which decays algebraically down to small
distances,
\be
v(d) \sim d^{-1}
\ee
as shown in the inset of Fig.~\ref{g1}.b. Such a functional dependence
derives from the form of HI at Oseen level~\cite{Llopis_jnnf}. As a
result of such coupling, two sedimenting filaments affect each other at
large distances, and the coupling becomes quantitatively relevant at
distances of the order of the filament's size.

For a given separation $d$, the velocity change due to hydrodynamic
cooperativity decreases with $B$. More rigid filaments have a larger
filament section exposed to the flow induced by the neighbour filament,
leading to a larger relative velocity increase.

%%%%%%%%%%%%%%%%%%%%%%%%%%%%%%%%
\subsection{Geometry II. Sedimenting Coplanar Filaments}\label{s6}
%%%%%%%%%%%%%%%%%%%%%%%%%%%%%%%%

\begin{figure}
\includegraphics[width=6.8cm, angle=-90]{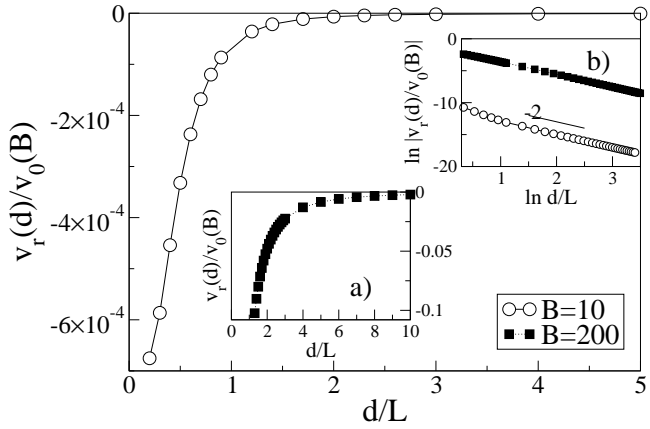}
\caption{Relative velocity of a pair of initially parallel semiflexible
filaments sedimenting on top each other (Geometry II) at $B=10$
normalized by the sedimenting velocity of a single filament at the
corresponding $B$, $v_0(B)$. a) Relative velocity  for $B=200$, where the
values of $v_r/v_0(B)$ are much larger. b) The algebraic dependence of
the relative velocity on the distance is compatible with $d^{-2}$.}  \label{vrel}
\end{figure}

We consider next a pair of straight parallel chains separated a distance
$d$ under the action of a uniform external field coplanar and transverse
to the two filaments, as depicted in Fig.~\ref{confi_ini_gen}.b; the
symmetry of the geometry ensures again $D=0$.
The upper chain bends less than the lower one, $A_1 < A_2$, and
sediments faster because it is subject to a smaller drag due to the
solvent counterflow.
Similar phenomena have been reported for the sedimentation of other
flexible objects, such as drops~\cite{mangastone,machu}. In
Fig.~\ref{vrel} we display the relative sedimentation velocity,
$v_r\equiv |v_{2z}|-|v_{1z}|$, as a function of a prescribed interfilament
distance, $d$.  The sedimentation velocity increases with $B$ until the
filament deformation reaches the plateau depicted in
Fig.~\ref{fig:onefilament}. The relative velocity vanishes at $B=0$ and
for larger values of $B$ increases up to a $10\%$. Accordingly, the
sedimentation velocity increases until the plateau regime of the
filament deformation is reached, where the time scales for displacement and bending do not differ significantly.  Since $v_r$ vanishes for $B=0$, the
relative velocity can change significantly upon increasing the
filaments' flexibility. Moreover, since the
relative velocity is always negative 
the two filaments will always approach and will eventually collide.
We have verified that the filaments sedimentation velocity decays as
$d^{-1}$, while their relative velocity, $v_r$, decays as $d^{-2}$
(Fig.~\ref{vrel}) because the leading contribution of $O(d^{-1})$
cancels out exactly. Such a behavior is general and valid for all values
of $B$ and in Appendix~\ref{ap_trim} we discuss such a dependence on the
basis of a simplified limiting model.

%%%%%%%%%%%%%%%%%%%%%%%%%%%%%%%%
\subsection{Geometry III. Collinear Filaments in a Transverse Field}\label{s7}
%%%%%%%%%%%%%%%%%%%%%%%%%%%%%%%%

Finally, we analyze the sedimentation of two collinear filaments under
the action of a uniform transverse field. To this end, we consider a
pair of filaments which are initially straight and with a minimal
bead-to-bead distance $d$, as shown in Fig.~\ref{confi_ini_gen}.c.

The hydrodynamic coupling induces a sedimentation velocity which
differs from free draining motion, $v_0$. Due to the instantaneous
propagation of HI in our model, deviatons from $v_0$ are observed
after one time step.  In Appendix~\cite{oseen_prediction} we compute this initial velocity when bending is negligible. 
The presence of the second filament induces in general a relative
displacement of the filaments and also a rotation because the mirror
symmetry is lost. The time scales at which filaments rotate and displace
depend on filament flexibility.

At short times, when filaments have deformed significantly although
their distance has not changed appreciably, one can characterize the
filaments by a sedimentation velocity, $v_s$, which can be understood
as the limit $v_s = \lim_{t \to 0}\ v(t) \neq v_0$. At long times,
filaments approach or move apart and rotate significantly only after
they have developed a well-defined bent shape.
 
\begin{figure}
\includegraphics[width=7cm,angle=-90]{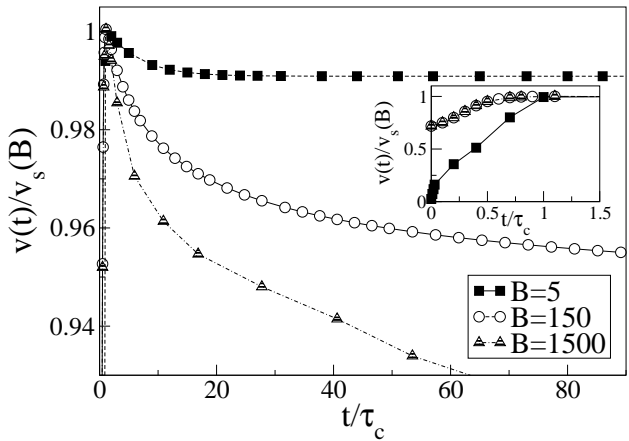} \caption{Sedimentation
velocity of a pair of filaments at $d/L=0.5$ as a function of time. At
$B=5$ the velocity reaches a plateau at small times which is
approximately stationary, whereas for {\bf $B=150$} the velocity
decays after the initial plateau at $t/\tau_c\sim1$. This dependence increases with $B$, as it becomes clear for the curve at $B=1500$.  Inset: velocity increase at short
times.}\label{vt}
\end{figure}

In Fig.~\ref{vt} we show the sedimentation velocity of a pair of
filaments at different values of B, as a function of time in units of
$\tau_c = L \xi_{\perp}/ F_e$, the time it takes a filament to displace
its own size, where $\xi_{\perp} = 4 \pi \eta L / \log(L/b)$ is the
friction coefficient of a rigid rod in the slender body limit.  One can
see how $v_s$ is reached on time scales of order $\tau_c$, and that a
second, smaller velocity is reached at larger times, when filaments
displace and $B$ is not too large. Upon increasing $B$, this second plateau becomes a slow decay toward the long-time regime. The  range of this decay also decreases with $B$.
  
In this geometry the long-time sedimentation regime is characterized
generically by a coupling between translation and rotation. We will
first describe the short-time sedimentation regime, and address the
long-time behavior subsequently.
\subsubsection{Short times}
\begin{figure*}[htp!]
\includegraphics[width=18.3cm,angle=0]{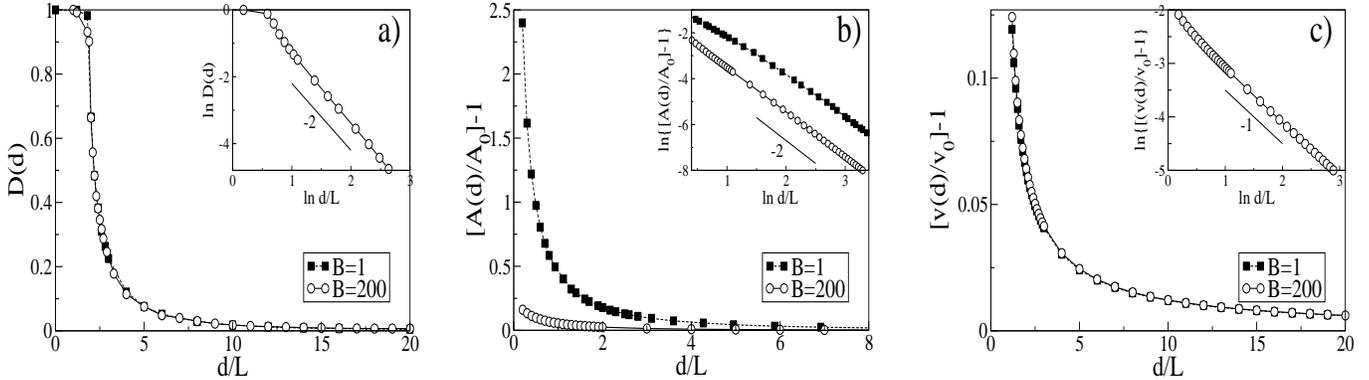}
\caption{Different measures of the configuration of a pair filaments
  in the short time regime at $B=1$ and $B=200$ for Geometry III. In
  all the insets we show the same curves in log-log scale to display
  their power law dependence which does not depend on $B$. a) Deformation asymmetry, $D(d)$. b) Bending
  amplitude, $A(d)$, normalized by the corresponding one filament
  value, $A_0$. c) Velocity in the direction of the external
  force. The asymptotic values are $v_0(B=1)=0.52 L/\tau_c$ and
  $v_0(B=200)=0.50 L/\tau_c$.}\label{Dav}
\end{figure*}

The presence of a neighbouring filament induces a deformation that is
not symmetric with respect to the center of mass of each filament. The
deformation asymmetry increases as the filaments approach, with an
algebraic dependence
%%%
\be
D(d) \sim d^{-2},
\ee
%%%
as displayed in Fig.~\ref{Dav}.a. The change in $D$ arises either
because of filament tilt, at small $B$, or due to bending, at large
$B$ values.
Analogous to the observations in the previous geometries, $A(d)$
decreases algebraically with filament separation, as shown in
Fig.~\ref{Dav}.b, although with an exponent $-2$ instead of $-3$.
The short-time sedimentation velocity $v_s(d)$ decays algebraically
toward the single filament value algebraically as $1/d$. However, the
dependence on $B$ is much weaker than in previous geometries, as
displayed in Fig.~\ref{Dav}.c. Thus, in the short-time regime we recover
practically a universal dependence of the velocity on the distance.
%%%%
\subsubsection{Long times}\label{long_times}
%%%%

In order to analyze the behavior at long times, it is useful to analyze separately the  small $B$ ($B<200$)and large $B$ ($B>200$) regimes, corresponding to filaments which do not reach and reach the saturation of single filament deformation, respectively, as shown in Fig.~\ref{fig:onefilament}).

\paragraph*{Small $B$}

\begin{figure}[htp!]
\includegraphics[width=7cm,angle=-90]{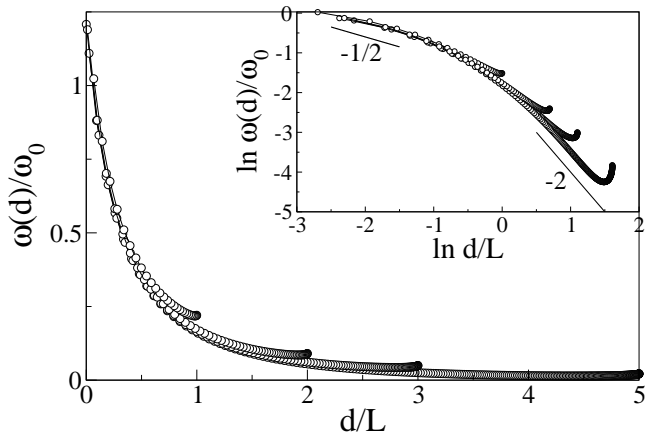}
\caption{Angular velocity, $\omega(d)$, of a pair of filaments as a
  function of the distance, for different initial separations, at
  $B=1$ for Geometry III. $\omega(d)$ is normalized by $\omega_0$, the angular
  velocity at the closest distance. The different curves
  correspond to different initial separations ($d(t=0) = 5, 3, 2, 1$ starting from right to left); after a transient
  related to the relaxation of the (initially straight) filaments, they
  follow a unique curve which depends only on the distance betwen the
  filaments. The inset shows the algebraic
  behavior at small and large distances. }\label{wd}
\end{figure}

\begin{figure*}[htp!]
\includegraphics[width=18cm,angle=0]{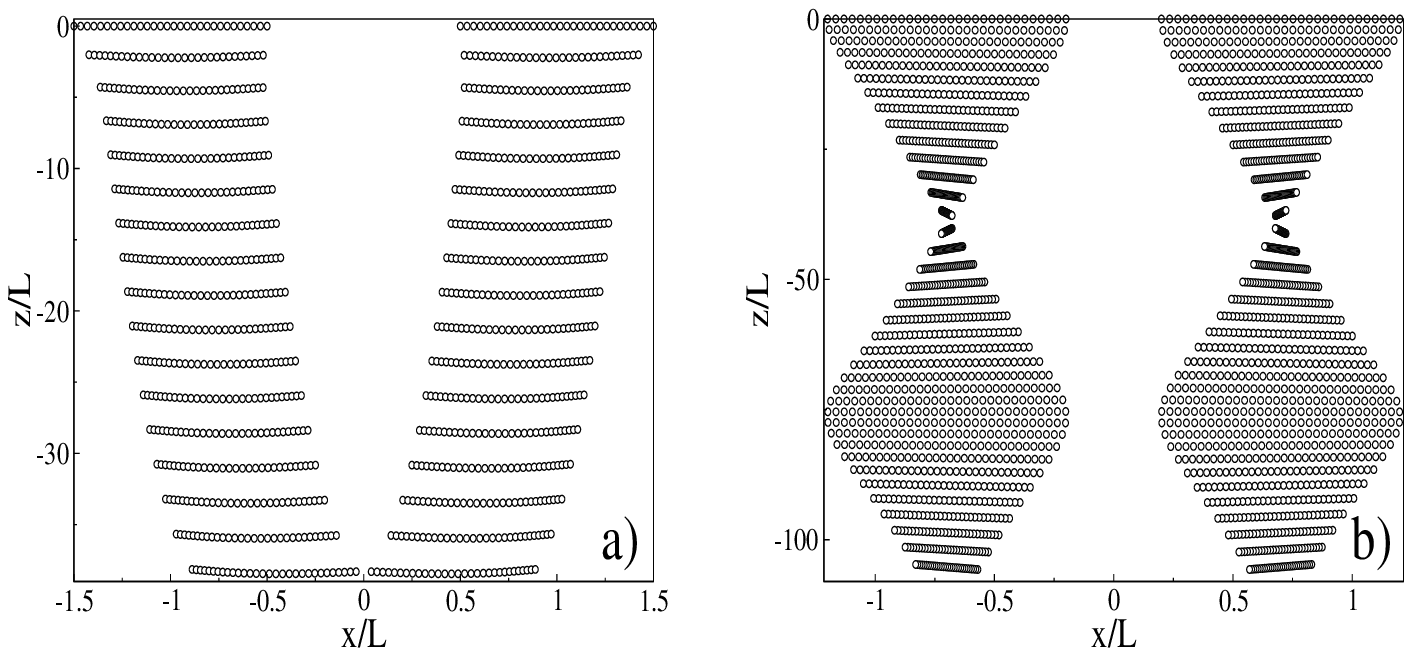}
\caption{a) Configurations of two initially straight filaments a distance  $d/L=1$ away from each other and  $B=100$ shown at intervals  $t/\tau_c=1.2$.  At long times ($t/\tau_c >> 50$) the two filaments eventually collide. b) Configurations of two initially straight filaments a distance  $d/L=0.4$ away from each other and  $B=1$ shown at intervals  $t/\tau_c=1.1$.  In this case we have fixed $d$  to highlight the
rotation of the filaments, characterized by a mean angular velocity which depends on the relative configuration of the two filaments. The apparent shortening of the filament is due to the
different scales in the two axis where distances are expressed  in units of  the filament length $L$. }\label{collcm}
\end{figure*}

In this regime,  two collinear parallel filaments always tilt, and  rotate as a result of the inhomogeneous hydrodynamic stress along  every filament.  As a result of the rotation and the geometrical asymmetry, we observe a relative velocity, $v_x$. In Fig.~\ref{wd} we display the angular velocity, $\omega$, which shows a crossover from a $d^{-2}$ dependence on filament separation to a
weaker $1/\sqrt{d}$ at shorter distances. On the contrary, $v_x$
decreases as $1/d$. The weaker dependence at short distances of the angular velocity develops as a result of the translation-rotation coupling; at a fixed distance the angular velocity decays always as $1/d^2$.  The weaker dependence on filament distance and small magnitude of
$\omega$ implies that filaments will usually approach and collide
before they have rotated by an angle larger than $\pi/2$, which would
allow them to move away from each other (see the example in
Fig.~\ref{collcm}.a).  Only initially remote filaments will avoid
collision on observable time scales.

Filament rotation can be clearly analyzed if the initial distance between the two filaments, $d$, is fixed. One can observe that the two filaments rotate with an averaged angular velocity which increases with decreasing $d$, as seen for example in Fig.~\ref{collcm}.b. In Appendix~\ref{rotation} we compute the  angular velocity for a pair of paralell sedimenting filaments, which shows the relevance of  the translation-rotation coupling.

\paragraph*{Large $B$}

\begin{figure*}[htp!]
\includegraphics[width=18cm, angle=0]{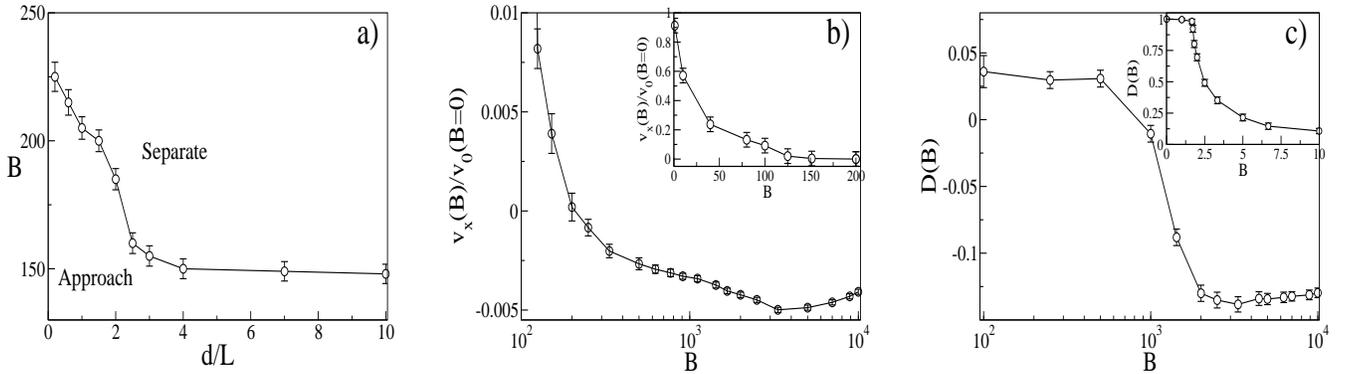}
\caption{Effective attraction and repulsion in geometry III.  a) $B-d$
  diagram showing the distinct behaviors of sedimenting collinear
  filaments as a function of initial distance and driving strength. b)
 Relative  velocity at $t=20\tau_c$, when $d=L$, as a function of
  $B$; positive velocity means attraction. Inset: decay of the relative velocity at small values of $B$. c) Deformation asymmetry at $d=L$ as a function of $B$,
  showing the change of sign and the minimum value at $B \sim
  3000$. Inset: decay at small values of $B$.}
\label{Bd}
\end{figure*}

\begin{figure*}[htp!]
\includegraphics[width=18cm,angle=0]{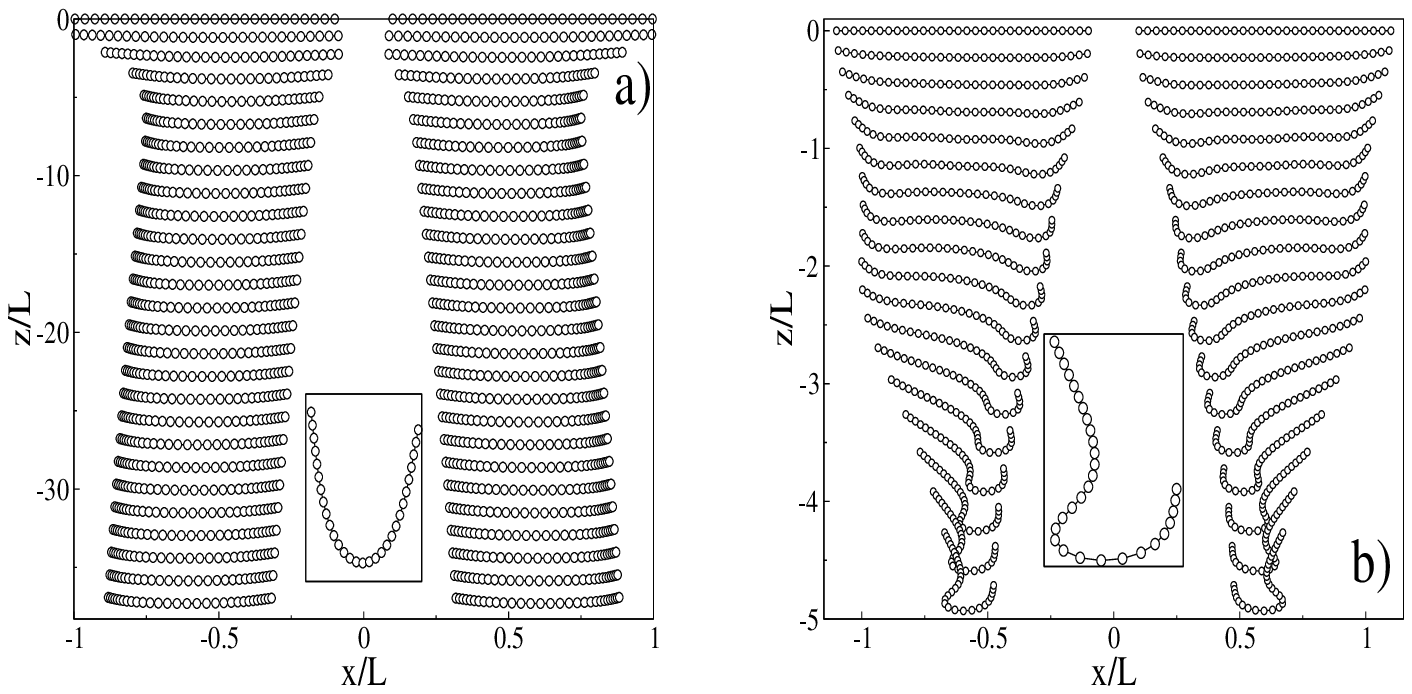}
\caption{a) Configurations of two initially straight filaments a distance  $d/L=0.2$ away from each other  and  $B=250$ shown at intervals  $t/\tau_c=2.0$.   Filaments repel each other and move apart with a highly curved shape, as shown in the inset for the left fiber. b) Configurations of two initially straight filaments a distance  $d/L=0.2$ away from each other  and  $B=10^4$ shown at intervals  $t/\tau_c=2.3+$.   Filaments initially evolve into the metastable configuration described in Fig.~\ref{fig:onefilament}. while they move apart from each other. During that separation the filaments  change their shape and recover transiently this metastable shape. In the inset we show one of the transient, deformed configurations the filaments explore.}
\label{conf_flexible}
\end{figure*}

When the degree of deformation is limited by the length of the
filament, its bending is less sensitive to the presence of a second
fiber. As a result, the angular velocity that characterizes rotation
decreases significantly with B. Associated to this reduced sensitivity, we observe that  for any degree of flexibility,
 at long times there exists a threshold $B^*(d)$, such as for $B < B^*$ the two
filaments approach, while for $B>B^{*}$  move apart after a transient induced by their initial condition,
as summarized in the $B-d$ diagram  in Fig.~\ref{Bd}.a; as $B$ increases
filaments move apart at smaller distances. Such a behavior is not present for rigid filaments, and it correlates with the component of
the velocity along the filament axis. In Fig.~\ref{Bd}.b we show such
a velocity as a function of $B$ for a given initial separation, on a
time scale in which filaments have displaced distances comparable to
their sizes. One can see that the velocity reverses sign at a finite
value of $B$, which in fact coincides with the change in behavior
displayed in the $B-d$ diagram.  When arriving at the bending plateau,
for $B>200$, filaments move apart (Fig.~\ref{conf_flexible}.a).  On
the same time scales we have computed the degree of asymmetry $D$, as
depicted in Fig.~\ref{Bd}.c. The behavior is qualitatively similar to
the one observed at short times.  As shown in the inset, for very small
values of $B$ the filaments essentially only rotate, and only  for $B>1$ flexibility starts to affect $D$ quantitatively. Such
parameter $D$ also reverses sign at a value of the flexibility similar
to that characterizing the change in the velocity. We attribute this
change in trend to a crossover from a regime of small $B$, where
filaments essentially rotate rigidly, to a regime where the asymmetric
deformation is controlled by bending.

At large values of $B$, a third regime is observed both in $v_x$ and
$D$; both quantities reach a minimum and decrease again in
magnitude. The minimum is observed in the parameter region where
metastable filament configurations for single filament sedimentation
develop.  At even larger values of $B$, we have also observed regimes
where intrafilament collisions are observed as a result of the shape
deformations the filament suffers during its sedimentation.
Therefore, the final configuration is not stable and changes
continuously with time.

%%%%%%%%%%%%%%%%%%%%%%%%%%%%%%%%
\section{Conclusions}\label{s8}
%%%%%%%%%%%%%%%%%%%%%%%%%%%%%%%%

We have studied the sedimentation of a pair of filaments suspended in
a low Reynolds number fluid.  The coupling that the filaments in the
solvent induce on each other through flows, the so called hydrodynamic
interactions (HI), give rise to a rich variety of cooperative
motion. We have concentrated on the simplest geometries, in order to
perform a careful analysis that allow us to focus on the essential
features of such cooperativity. To this end, we have implemented and
used a simple and efficient numerical method which models the filament
as a set of beads and imposes inextensibility where HI are treated at
the Oseen approximation.

The geometries we considered have helped us to show that in all cases
the sedimentation of a pair of semiflexible filaments is qualitatively
different from that of rigid filaments. The rigid limit is in fact
singular, and sets in because the time the filaments need to modify
their initial configuration increases with filament rigidity, and
diverges for infinitely rigid rods. For sufficiently symmetric
geometries, such as Geommetry I, HI modify the degree of deformation
of each filament and its final sedimentation velocity. The interaction
decays algebraically with filament distance, and it becomes
quantitatively relevant for separations of the order of the filaments'
size.

For sedimenting parallel coplanar filaments, we have shown that the top 
filament  bends more and moves
faster, inducing pair collision, as opposed to the sedimentation of
rigid rods. 

In less symetric situations, the hydrodynamic coupling induces a
rotation and translation of the sedimenting filaments, because their
bending lacks the symmetry with respect to the filaments' center of
mass.  In these cases, pairs of filaments still interact at long
distances, but their sedimentation behavior becomes more involved, and
depends on their degree of flexibility as well as their initial
conditions.  We have shown that such rich behavior includes periodic
bound trajectories, filament rotation as well as sedimentation with
unsteady conformations. 

To sum up, filament flexibility and hidrodynamic coupling modify
profoundly the behavior of filament sedimentation; the simplified
geometries explored have helped to understand the interplay between
elasticity and hydrodynamics and opens the possibilties to analyze in
detail how such interactions modify the response of filament
suspensions to applied external fields. The study we have carried out
has allowed us to make definite predictions that can be tested in
controlled experiments, for example with a centrifuge coupled to
optical microscopy.

\begin{acknowledgments}
We are grateful to H.A.~Stone, for pointing out to us the interest of
geometry II. I.Ll. and I.P. acknowledge financial support from DGCIYT of
the Spanish Government (FIS2005-01299). I.P. thanks {\sl Distinci\'o de
la Generalitat de Catalunya } for financial support.
\end{acknowledgments}
%%%%%%%%%%%%%%%%%%%%%%%%%%%%%
%%%%%%%%%%%%%%%%%%%%%%%%%%%%%
\appendix
%%%%%%%%%%%%%%%%%%%%%%%%%%%%%
%%%%%%%%%%%%%%%%%%%%%%%%%%%%%
%%%%%%%%%%%%%%%%%%%%%%%%%%%%%%%%
\section{Initial sedimentation velocities}\label{oseen_prediction}
%%%%%%%%%%%%%%%%%%%%%%%%%%%%%%%%

In the initial stages of their sedimentation, straight filaments have
not deformed significantly. In this regime it is possible to obtain
analytical expressions for their sedimentation velocities at the Oseen
level.

In particular, we are interested in the velocity that a straight
filament oriented along the $x$ direction induces in a second
collinear filament a distance $d$ away, as depicted in
Fig.~\ref{confi_ini_gen}.c. The external field is applied
perpendicular to both filaments, along the $z$ direction, and hence
the distance between beads reduces to their separation along the $x$
direction. As a result, the velocity on bead $i$ due to bead $j$ in
the direction of the external force, can be expressed as \be v_{i1}^H
= \frac{3 a}{4} \frac{F_e}{\gamma_0} \sum_j \frac{1}{|x_j - x_i|},
\label{a1} \ee
where we have assumed that bead $i$ belongs to the filament on the
left while $j$ is a filament belonging to the filament on the right
hand side of the pair, and hence the sum runs over the beads of this
second filament. If we approximate the sum by an integral over the
length of the filament, we arrive at
\be
v_{i1}^H = \frac{3 a}{4 L} \frac{F_e}{\gamma_0} \int_{L+d}^{2L+d} \frac{\,dx_j}{|x_j - x_i|} =
\frac{3 a}{4 L} \frac{F_e}{\gamma_0} \ln \left(\frac{2L + d - x_i}{L+d - x_i} \right).
\ee
The contribution of the second filament to the sedimentation velocity
of the first one is obtained by computing the  induced center of mass velocity. In the continuum
approximation, this induced velocity can be written down as $v_1^H =
1/L \int_0^L v_{i1}^H \,dx_i$, leading to
\be
v_1^H = \frac{3 a}{4 L} \frac{F_e}{\gamma_0} \left\{ \frac{d}{L} \ln
\left[\frac{d(d+2L)}{(d+L)^2}\right] + 2 \ln \left[ \frac{d+2L}{d+L} \right]
\right\}.
\ee
We can proceed analogously, with obvious modifications in the
geometry, to obtain the contribution of a second filament to the
sedimentation velocity of the reference one for two coplanar filaments (geometry II). In this case we arrive at 
\be
v_1^H = \frac{3 a}{4 L} \frac{F_e}{\gamma_0} \left\{2 \sqrt{\frac{d^2 + L^2}{L^2}}  - 2 \frac{d}{L} + \ln \left[ \frac{L +
\sqrt{d^2 + L^2}}{-L + \sqrt{d^2 + L^2}} \right] \right\}.  
\ee
Finally, for Geometry I, if we take the axis of the filaments along
the $x$ axis and their distance along the $z$ we get, 
\be 
v_1^H =
\frac{3 a}{4 L} \frac{F_e}{\gamma_0} \left\{ 2 \frac{d}{L} - 
2 \sqrt{\frac{d^2 + L^2}{L^2}} + \ln \left[ \frac{L + \sqrt{d^2
+ L^2}}{-L + \sqrt{d^2 + L^2}} \right] \right\}.  \label{a5} 
\ee 
In geometries I and II the velocity diverges as the two filaments
approach each other because all the beads become infinitely close, as opposed to geometry III which is characterized by a finite induced sedimentation velocity at contact, $3 \ln(\sqrt{2}) a F_e/L\gamma_0$.

%%%%%%%%%%%%%%%%%%%%%%%%%%%%%%%%
\section{Trimers in Geometry II}\label{ap_trim}
%%%%%%%%%%%%%%%%%%%%%%%%%%%%%%%%

In Appendix~\ref{oseen_prediction}, we have computed the initial
sedimentation velocity for a pair of filaments where one of them moves
on the wake of the neighbouring one (eq.~\ref{a5}). Here we provide an
estimate of the rate at which they approach each other when filament
deformation is small (either because we focus at short times, or
because $B$ is small). We consider the simplest possible case, where
the filaments are represented by trimers.

\begin{figure}
\includegraphics[width=7.5cm,angle=-90]{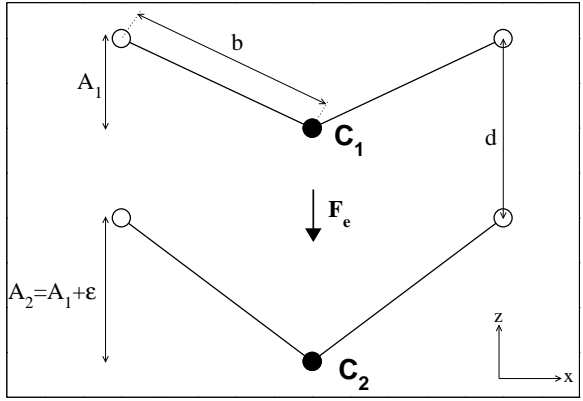} \caption{Configuration
of a pair of trimers under the action of an external force. The
difference on bending amplitude is given by the parameter $\epsilon$.}
\label{trimer}
\end{figure}

We will calculate the hydrodynamic velocity due to the beads of the
neighbor filament on the central beads (depicted in black in
Fig~\ref{trimer}), defined as $c_1$ and $c_2$ in Fig~\ref{trimer}.  The
difference between the two velocities is a measure of their relative
velocity.

We characterize $d$ as the distance between the corner beads and the
bending amplitude as the separation between the central and corner beads
of each filament along the direction of the applied field. The top and
bottom chains will bend an amount $A_1$ and $A_2$ respectively, where
$A_2 \equiv A_1+\epsilon$. Hence, $\epsilon$ measures the relative
bending amplitude due to the different hydrodynamic coupling. The
distance between consecutive beads of a given trimer is $b \equiv L/2$,
and the force field has a magnitude $F_e$.

The velocity on the central beads $1$ and $2$ has only component in
the $z$ direction,
\be 
v_{c1} = \frac{3 a}{2}
\frac{F_e}{\gamma_0} \left\{ \frac{2(d-A_1)^2 + b^2
-(A_1+\epsilon)^2} {\left[(d-A_1)^2 + b^2 -(A_1+\epsilon)^2
\right]^{3/2}} + \frac{1}{d+\epsilon} \right\} 
\ee

\be
v_{c2} = \frac{3 a}{2}
\frac{F_e}{\gamma_0} \left\{ \frac{2(d+A_1+\epsilon)^2 + b^2
  -A_1^2}{\left[(d+A_1+\epsilon)^2 + b^2 -A_1^2 \right]^{3/2}} +
\frac{1}{d+\epsilon} \right\}.
\ee
Therefore, the relative velocity, $v_r \equiv v_{c1} - v_{c2}$, can be expressed as,
\begin{eqnarray}
v_r & = & \frac{3 a}{2} \frac{F_e}{\gamma_0} \left\{ \frac{2(d-A_1)^2 + b^2
  -(A_1+\epsilon)^2}{\left[(d-A_1)^2 + b^2 -(A_1+\epsilon)^2
    \right]^{3/2}} \right. \nonumber \\ & - &
\left. \frac{2(d+A_1+\epsilon)^2 + b^2
  -A_1^2}{\left[(d+A_1+\epsilon)^2 + b^2 -A_1^2 \right]^{3/2}}
\right\}.
\end{eqnarray}
For small $B$, or large distances, also the differential bending will
be small. In this case, if we expand the previous expression in powers
of $\epsilon$, we arrive at 
\be 
v_r (d, \epsilon) = \frac{3 a
F_e}{\gamma_0} \left(2A_1+\epsilon+\epsilon^2\right) \frac{1}{d^2} +
O(\epsilon^3,d^{-4}) \ ,  
\ee 
which we have validated using simulations of trimers in this geometry. This
expression shows that the top trimer moves faster as a result of
HI. The relative velocity depends on the distance as $1/d^2$.  The increase of $v_r$ with $A_1$
indicates that also the relative velocity will increase with the
flexibility $B$.

%%%%%%%%%%%%%%%%%%%%%%%%%%%%%%%%
\section{Rotation of filaments}\label{rotation}
%%%%%%%%%%%%%%%%%%%%%%%%%%%%%%%%
In section~\ref{long_times} we have seen that in geometry III, small B
filaments rotate and that the parallel filament geometry is
unstable. In this appendix we consider a pair of straight filaments,
with vanishing bending and constraint forces, oriented along the
direction of the external field, $F_e$, separated a distance $d$ on
the $x$ direction. Hence, we can compute the velocity on each bead due
to the presence of the second filament following the same approach as
in Appendix~\ref{oseen_prediction}. The components of the velocity on
monomer $i$ of a given filament in the directions along and
perpendicular to the filaments can be expressed as
 
\begin{eqnarray}
\left( v_i^H \right)_x &=& \frac{3 a}{4 b} \frac{F_e}{\gamma_0} \sum_{j \neq
  i} \frac{ x_{ij}
z_{ij} }{ r_{ij}^3 /b},\nonumber\\
\left( v_i^H \right)_z &=& \frac{3 a}{4 b} \frac{F_e}{\gamma_0} \sum_{j \neq
  i} \left[\frac{1 + (z_{ij}/r_{ij})^2 }{ r_{ij}/b } \right] ,
\end{eqnarray}
where $x_{ij} = |x_i - x_j| \equiv d$ , $z_{ij} = |z_i - z_j|$ and
$r_{ij} = \sqrt{x_{ij}^2 + z_{ij}^2}$, and where the sums run over the
monomers of the same and the neighbouring filaments. Due to the
symmetry of the configuration, $(v_i^H)_x$ only has
nonvanishing contributions from the neighbouring filament, while
$(v_i^H)_z$ has a contribution for the filament itself, which corresponds to the
sedimentation velocity of an isolated filament aligned parallel to the
external field, and which in the continuum approximation leads to a
sedimentation velocity $(v_1^H)^{\infty}_z= F_e \ln(L/b)
/ 2 \pi \eta L$ in the slender body limit. The contribution of the
neighbouring filament increases this velocity to,

\be
\left( v_1^H \right)_z=  \left( v_1^H \right)^{\infty}_z + \frac{3 a}{4 L}
\frac{F_e}{\gamma_0} 
\left[\frac{2}{3 d} + \frac{2}{L} \ln(L+\sqrt{d^2+L^2})\right].   
\label{totalvel}
\ee 

While the filament sediments, it will experience a transverse
velocity. Due to the symmetry, this velocity distribution does not
lead to any transverse translation; in fact, $(v_1^H)_x=(1/L)\sum_j (v_j^H)_x=0$. This velocity profile
yields a net rotation of the filament, which is given, at large
distances by $I \omega = m \sum_{ij} z_{ij} (v_i^H)_x= (3 a m F_e/4
\gamma_0 L) \left[3/d - 7 L^2/12 d^4\right]$, where $I= m L^2/12$ is
the moment of inertia, hence the angular velocity decays
asymptotically as $1/d^2$, i.e. faster than the approaching velocity $
(v_1^H)_z$, \be \omega = \frac{9 a F_e}{4 \gamma_0 L^2} \left[
3\frac{1}{d^2} - \frac{7 L^2}{12} \frac{1}{d^4} \right].
\label{hydrovel_ang}
\ee
%%%%%%% Bibliografia %%%%%%%
 
\end{document}